\newlength\smallfigwidth
\newlength\figwidth
  \documentclass[aps,twocolumn,floatfix,amsmath,amssymb,showkeys,letterpaper]{revtex4}

\newcommand{\be}{\begin{equation}}
\newcommand{\ee}{\end{equation}}
\newcommand{\sech}{\, \text{sech}}

\newcommand{\ba}{\begin{align}}
\newcommand{\ea}{\end{align}}
\newcommand{\bn}{\begin{eqnarray}}
\newcommand{\en}{\end{eqnarray}}

\newcommand{\bsub}{\begin{subequations}}
\newcommand{\esub}{\end{subequations}}

\usepackage{graphicx}
\usepackage{amsmath}
\usepackage{hyperref}
\usepackage{bm}
\usepackage{color}
\definecolor{chcolor}{rgb}{0.8,0.1,0.1}

\usepackage{soul}

\begin{document}

\noindent
\title{Domain walls in a dipole-coupled transverse magnetic island chain}
\author{G. M. Wysin}
\email{wysin@k-state.edu}
\homepage{http://www.phys.ksu.edu/personal/wysin}
\affiliation{Department of Physics, Kansas State University, Manhattan, KS 66506-2601}

\date{Mar. 17, 2026}
\vskip 0.2in
\begin{abstract}
I analyze the nonlinear Hamiltonian equations of motion for a one-dimensional chain of transverse magnetic
nano-islands, seeking solutions for different types of static domain-walls (DWs) connecting uniform static states.
The system of elongated magnetic islands oriented transverse ($y$-direction) to the chain direction ($x$-direction)
experiences an applied magnetic field transverse to the chain.  The macro-spin model includes dipole interactions between 
islands, their uniaxial and easy-plane anisotropies, and Oersted energy of the applied field. DWs can form most 
easily between pairs of degenerate uniform states, described by their local magnetizations as oblique, $y$-parallel, 
and $y$-alternating. The DWs between oblique states are well-described with scalar $\varphi^4$ theory. 
General DW structures are found via a numerical energy relaxation scheme.  At some anisotropy and field parameters, 
nearest-neighbor dipole interactions drive antiferromagnetic order inside the DW itself.  The variety of DWs 
present in the model might be exploited for their sensitivity to parameter changes in detectors or 
switching technology.
\end{abstract}

\maketitle

\section{Introduction: Multi-stable magnetic island chains}
Arrays of elongated magnetic elements on a nonmagnetic substrate such as two-dimensional (2D) artificial spin 
ice \cite{Wang06,Nisoli13,Nguyen17,Skjaervo20} and one-dimensional (1D) dipolar chains 
\cite{Wittborn+99,Ostman18,Zhang+18}
can exhibit many interesting features due to the frustrated competition between geometric anisotropies and 
dipolar interactions.  Whether 1D or 2D, the possible uniform and degenerate allowed states with differing 
magnetizations can be expected to connect to each other through domain walls (DWs). While 2D systems have an
enormous variety of designs and allowed states, 1D systems offer a more direct analysis of a limited number of
uniform states and likely DW configurations. The studies here concern engineered magnetic islands like those
in spin ice (Permalloy or other media \cite{Garcia+02a}), but along a 1D chain. The results could be applicable 
more generally to other 1D chains, such as Fe nanoparticles \cite{Tang+15}, nanowire elements \cite{Garcia+02b}, 
or possibly Co$_2$C nanoparticles \cite{Zhang+18} and biomineralized magnetosomes \cite{Wittborn+99}.
 
In earlier work a model for a linear chain of elongated magnetic islands was considered, where the thin 
islands are oriented with their longer axes perpendicular to the chain direction \cite{Wysin22,Wysin24}.  
The chain direction defines an $x$-axis and the transverse direction is labeled the $y$-axis.  The model 
bears some resemblance to a 1D artificial spin-ice.  Each island has shape anisotropy that 
tends to make its magnetic dipole point along the island's longer axis, i.e., easy-axis anisotropy of strength 
$K_1$.  This competes, however, with the dipolar interactions among the dipoles, of strength $D$ between nearest 
neighbors, and the system is frustrated.  For strongly elongated islands, the ground state will consist
of the dipoles alternately pointing along $+y$ and $-y$, which mimics the ground state in square lattice 
artificial spin-ice.

\begin{figure}
\includegraphics[width=\figwidth,angle=0]{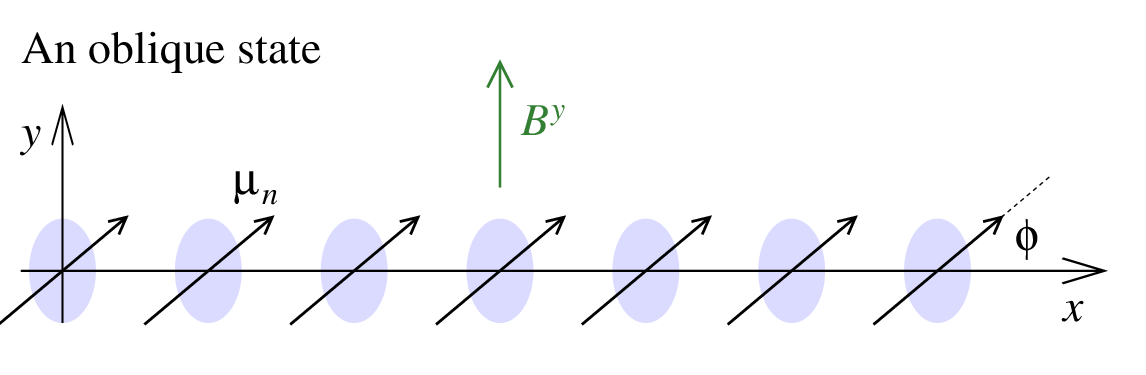}
\caption{\label{oval-islands-oblique} Part of a system of magnetic islands elongated transverse to the chain,
with the dipoles in a uniform oblique state due to the competition among dipolar energy, shape anisotropy, and
Oersted energy.}
\end{figure}

If there is no applied field, the strength of the easy-axis anisotropy relative to the dipole interactions controls 
the allowed uniform states.  At very weak anisotropy, 1D dipolar interactions for any distance dominate and all the dipoles 
point along the chain direction, in an $x$-parallel state.  At intermediate anisotropy, the nearest-neighbor dipole interactions
dominate and the dipoles alternately point transversely along $+y$ and $-y$ in a $y$-alternating state ($y$-alt for short).  
At strong anisotropy, the $y$-alt states are still the lowest energy states, but it is also possible for the dipoles to point 
all along $+y$ or $-y$ in metastable $y$-parallel or transverse states. 

A transverse magnetic field $B^y$ can change the states and affect their stability. The dipoles in an $x$-parallel state tilt
towards the field direction--they are transformed to oblique states with magnetization pointing at an oblique angle
to the chain direction, as in Fig.\ \ref{oval-islands-oblique}.  A transverse magnetic field even helps to stabilize 
oblique states for intermediate anisotropy strength, where the $x$-parallel states were prohibited for zero magnetic field.  
At strong magnetic field, all dipoles will align to the magnetic field in a $y$-parallel or transverse state.  Generally 
the stability regions for the oblique, $y$-parallel, and $y$-alt states have been mapped out in terms of the parameters 
$K_1/D$ and $\mu B^y/D$, where $\mu$ is the islands' magnetic dipole moment \cite{Wysin24}. 

Oblique states are doubly degenerate: their magnetization tilts towards $B^y$ with either a positive or negative component
along $x$.   The $y$-alt states are also doubly degenerate: either the even lattice sites or odd lattice sites point
towards $y$, but their energy is unaffected by the applied field.  Then the question arises,  what is the structure of
domain walls that certainly must be possible, connecting these degenerate states?  That is, what is the structure of
a DW between the two different oblique states, or between the two $y$-alt states?  Further, can there be domain walls
even between non-degenerate states, such as between the minimum energy configuration and a metastable configuration? 
An example of the latter would be DW between an oblique state and a $y$-alt state, for parameters where they can coexist.  

To address these questions, a local energy minimization algorithm is used to relax a system into a possibly smooth DW configuration,
starting from initial conditions where the two halves of the chain are in different uniform states.  The energy minimization
is accomplished by iteratively pointing each dipole along the direction of the effective magnetic field that acts on it. 
Eventually all dipoles point parallel to their respective effective fields, which is a stable situation of zero torque 
on each dipole.  Such a configuration will be a local energy minimum, but it should be stable against small fluctuations.  
Most of this study includes all of the dipole pair interactions in the system, denoted as the long-range dipole (LRD) model.
For comparison, it is fruitful to consider a fictitious nearest-neighbor (NN) model, limiting dipole interactions to first
neighbors.  The NN model lends itself to an approximate analytic solution for DWs connecting oblique states, and that analysis 
is partly successful when extended to the LRD model.

\section{The system with transversely oriented magnetic islands}

For single-domain magnetic islands (small enough with strong internal ferromagnetic exchange),
the state of one island can be approximated as a single magnetic dipole of fixed magnitude $\mu$,
and direction vector ${\bf S}_n$ (macro-spin approximation) \cite{Sayad+12,Wysin+13}.
A particular dipole is denoted $\vec\mu_n = \mu {\bf S}_n$, where ${\bf S}_n$ is a unit macro-spin vector.
It will be convenient to write these spin vectors using planar spherical angles ($\phi_n,\theta_n)$,
where $\phi_n$ is an azimuthal angle in the $xy$-plane and $\theta_n$ is the tilting of a
spin out of the $xy$-plane, i.e.,
\be
\label{coords}
{\bf S}_n=(\cos\theta_n \cos\phi_n, \cos\theta_n \sin\phi_n, \sin\theta_n).
\ee
For classical mechanics, $S_n^z=\sin\theta_n$ is the momentum conjugate to generalized coordinate $\phi_n$.
The islands have an easy-plane anisotropy that tends to keep their dipoles in the $xy$-plane,
while shape anisotropy leads to an easy-axis anisotropy along their longer axes, which are perpendicular
to the chain direction.  The dipoles also interact with a magnetic field applied transverse to the chain 
direction.

\subsection{Hamiltonian with anisotropies, applied field and dipolar interactions}
Pairs of dipoles on a one-dimensional (1D) grid interact via long-range dipolar interactions.
It is useful to define the energy constant for the interaction of nearest-neighbor (NN) pairs,
\be
D = \frac{\mu_0 \mu^2}{4\pi\; a^3},
\ee
where $a$ is the center-to-center spacing between
neighboring islands and $\mu_0$ is the magnetic permeability of vacuum.  
The dipole interaction is reduced relative to this by the cube of the
distance $r$ measured in lattice constants, which is an integer $k=r/a$.
Then the Hamiltonian for a chain of $N$ dipoles with anisotropies and exposed to a uniform transverse magnetic field 
of strength $B^y$ along the $y$-direction is
\begin{align}
\label{Ham}
H & = \sum_{n=1}^{N}  \Big\{ \sum_{k=1}^{R} \frac{D}{k^3}
\left[ {\bf S}_n\cdot {\bf S}_{n+k} -3({\bf S}_n\cdot \hat{x})({\bf S}_{n+k}\cdot \hat{x})\right] 
\nonumber \\
& -K_1 \left(S_n^y\right)^2 +K_3 \left(S_n^z\right)^2 -\mu B^y S_n^y \Big\}.
\end{align}
Parameters $K_1$ and $K_3$ represent easy-axis and easy-plane anisotropies.  The islands have their longer axes 
along the $y$-direction. Shape anisotropy makes the dipoles energetically prefer to point along $\pm y$-directions  
by energy $K_1$ over the $\pm x$-directions.  The islands also are considered thin in the $z$-direction, which helps to 
restrict their dipoles to stay in the $xy$-plane by an energetic preference of $K_3$. 
The last term is the Oersted interaction of the dipoles with the transverse applied field $B^y$.

An upper limit $R$ is used on the range of the dipolar interactions. Note that each dipole pair interaction
is included only once.  When $R=1$, $H$ reverts to a NN model. 
For theory it is useful also to consider $R\to\infty$, which is referred to as the long-range-dipole (LRD) model. 
However, most of the numerical simulations obtain the important results with systems of $N=200$ islands,
in which effectively $R=200$, but with end effects. Obviously the sum over $k$ is cut off if site $n+k$ falls
outside the finite system.  No attempt is made to eliminate end effects, which would be expected as real physical 
effects in experiments.   

In terms of the spin-angles,  one gets an alternative useful expression of the Hamiltonian,
\begin{align}
\label{Hphitheta}
& H  = \sum_{n=1}^{N} \Big\{ \sum_{k=1}^R \frac{D}{k^3} \big[\sin\theta_n\sin\theta_{n+k}  \\
&   +\cos\theta_n \cos\theta_{n+k} 
 (-2\cos\phi_n\cos\phi_{n+k}+\sin\phi_n\sin\phi_{n+k})\big] \nonumber \\
&   -K_1\cos^2\theta_n \sin^2\phi_n +K_3\sin^2\theta_n -\mu B^y \cos\theta_n\sin\phi_n \Big\}. \nonumber
\end{align}
For discussion of domain walls,  I consider a long enough system that the ends remain in different 
nearly uniform states.  Although one may think some special boundary conditions are needed to hold the 
opposite ends of the chain in different configurations, that is not required.  The simulations done here
have free or open boundary conditions. There is simply an abrupt end to the system beyond which no more
sites are present.  The almost-uniform configurations outside of the desired domain wall remain there due
to their local energetic stability.  They are either stable (the lowest available uniform state) or 
metastable (not the lowest, but a local energy minimum). In most cases, the two boundary states even have
the same energy per island.

\subsection{Spin Dynamics and Effective Magnetic Fields}
I appeal to Hamiltonian spin dynamics to define the effective magnetic field that acts on each dipole.
Suppose there is a gyromagnetic ratio $\gamma_{\rm e}$ that converts angular momenta ${\bf L}_n$ into magnetic
dipoles via $\vec\mu_n = \gamma_{\rm e}{\bf L}_n$.  Then the undamped free dynamics of a magnetic dipole system
follows a torque equation (see Ref.\ \cite{Wysin15}, Ch.\ 5),
\be
\label{dynam}
\frac{d{\bf L}_n}{dt} = \frac{1}{\gamma_{\rm e}} \frac{d\vec\mu_n}{dt} = \vec\mu_n\times {\bf F}_n ,
\ee
where the Hamiltonian can be written with effective magnetic fields ${\bf F}_n$ at each site,

\be
H=-\sum_n \vec\mu_n \cdot {\bf F}_n.
\ee
Those are seen to be obtained from $H$ by derivative operations for each component,
using $\vec\mu_n = \mu {\bf S}_n$,
\be
\label{eff-B}
{\bf F}_n = -\frac{\partial H}{\partial \vec\mu_n}
= \frac{-1}{\mu} \left( \frac{\partial H}{\partial S_n^x}, \frac{\partial H}{\partial S_n^y}, \frac{\partial H}{\partial S_n^z}  \right).
\ee
The Hamiltonian (\ref{Ham}) produces the following expressions
for the Cartesian components of the effective fields, 
\begin{align}
\label{Fnxyz}
F_n^x & = 
-\frac{1}{\mu}\sum_{k=1}^R \frac{-2D}{k^3}  \left( S_{n+k}^x +S_{n-k}^x \right),
\nonumber \\
F_n^y & = 
-\frac{1}{\mu}\left[ \sum_{k=1}^R \frac{D}{k^3} \left( S_{n+k}^y +S_{n-k}^y \right) -2 K_1 S_n^y -\mu B^y \right],
\nonumber \\
F_n^z & = 
-\frac{1}{\mu}\left[ \sum_{k=1}^R \frac{D}{k^3}  \left( S_{n+k}^z +S_{n-k}^z \right) +2K_3 S_n^z \right].
\end{align}
These simple expressions are efficiently evaluated by computer.  In a finite chain, the dipolar sums will terminate 
at the chain ends, for open-ended boundaries.

For theory, it is better to use angular coordinates, because there are only two.
Transforming the dipoles to the spherical coordinates in (\ref{coords}), the mechanics is that
where $\phi_n$ are generalized coordinates and $\sin\theta_n$ are the corresponding conjugate
momenta.  The dynamics obeys Hamiltonian equations,
\begin{align}
\label{HE}
\frac{\mu}{\gamma_{\rm e}} \frac{d}{dt}\phi_n = \frac{\partial H}{\partial \sin\theta_n}, 
\qquad
\frac{\mu}{\gamma_{\rm e}} \frac{d}{dt}\sin\theta_n = -\frac{\partial H}{\partial \phi_n}.
\end{align}
These dynamic equations apply to any situation, regardless of the boundary states
outside of the domain wall that is being found. To apply these, it is convenient to
define a rescaled time variable,
\be
\tau \equiv \frac{\gamma_{\rm e}}{\mu} t,
\ee
and generally use dot on top of a variable to indicate the derivative, $d/d\tau$.

\subsection{Static configurations from energy minimization of a discrete chain}
\label{sims}
For numerical solutions, especially static ones, energy minimization can be used to get some soliton-like
or topological solutions.   I want to find configurations that connect two of the uniform states.
Those will set the boundary conditions at $x=\pm\infty$. In practice, I initially set each half of the finite 
system in one of the two uniform states, with an abrupt connection between them. Then a very small amount of 
randomness was applied to the initial dipoles to break any perfect symmetry, for checking the robustness of 
the final state.
 
The energy minimization can be done from a physical standpoint by iteratively setting each dipole to point along 
the effective magnetic field ${\bf F}_n$ caused by the rest of the system, as given in Eqs.\ (\ref{eff-B}) and (\ref{Fnxyz}). 
Then the iteration involves scanning through the system and setting the dipole at site $n$ to align along the direction
of ${\bf F}_n$, i.e., a replacement
\be
{\bf S}_n = \frac{{\bf F}_n}{\left\vert {\bf F}_n \right\vert}.
\ee
This may indeed overshoot beyond the required solution, and there may be ways to modify it so the process does not
become unstable or oscillate between some solutions.  Usually this algorithm usually works well and the process can 
be stopped when the relative changes in the ${\bf S}_n$ components become less than some desired value (on the order of $10^{-8}$).
By then, the energy is not changing.

If the iteration overshoots or oscillates,  it can be augmented by adding in a Lifshitz-Gilbert damping field with
parameter $\alpha < 1$ into the time-derivative of the equation of motion. Then Eq.\ (\ref{dynam}) is modified as  
\be
\label{dynam1}
\frac{d{\bf L}_n}{dt} = \frac{1}{\gamma_{\rm e}} \frac{d\vec\mu_n}{dt} 
= \vec\mu_n\times \left[ {\bf F}_n -\alpha \left( \vec\mu_n \times {\bf F}_n \right) \right] .
\ee
In this case, let the net effective field be
\be
{\bf N}_n = {\bf F}_n -\alpha \left( \vec\mu_n \times {\bf F}_n \right). 
\ee
Then the iteration proceeds by aligning the dipoles to this new effective field,
\be
{\bf S}_n = \frac{{\bf N}_n}{\left\vert {\bf N}_n \right\vert}.
\ee
Generally, this may change the rate at which the system moves downhill in total energy, but for this simple
1D problem it does not seem to affect the final state, unless there are many different possibilities
with similar energies or with high frustration. I used this procedure with $\alpha$ from 0.01 to 0.1 to 
find domain walls easily and quickly.

\subsection{The angular equations of motion on a lattice}
For theory it is good to have the dynamic equations (\ref{HE}) in terms of the spin angles. 
Finding the needed derivatives with respect to $\sin\theta_n$, 
the in-plane angle follows
\begin{align}
\label{dotphi}
& \dot\phi_n  =\frac{\partial H}{\partial \sin\theta_n} =   
\sum_{k=1}^R \frac{D}{k^3} \Big[ \sin\theta_{n-k}+\sin\theta_{n+k}   \\
& -\tan\theta_n \cos\theta_{n+k}(-2\cos\phi_n\cos\phi_{n+k}+\sin\phi_n\sin\phi_{n+k}) 
\nonumber \\
& -\tan\theta_n \cos\theta_{n-k}(-2\cos\phi_n\cos\phi_{n-k}+\sin\phi_n\sin\phi_{n-k}) \Big]
\nonumber \\
& +2K_1 \sin\theta_n \sin^2\phi_n +2 K_3 \sin\theta_n +\mu B^y \tan\theta_n \sin\phi_n . \nonumber
\end{align}
The derivatives of $H$ with respect to $\phi_n$ lead to:
\begin{align}
-\dot\theta_n = & \frac{1}{\cos\theta_n} \frac{\partial H}{\partial \phi_n} =  
 \sum_{k=1}^R \frac{D}{k^3} \Big[ \nonumber \\
&  \cos\theta_{n-k} (2\sin\phi_n \cos\phi_{n-k} +\cos\phi_n \sin\phi_{n-k})
\nonumber \\
+ & \cos\theta_{n+k} (2\sin\phi_n \cos\phi_{n+k} +\cos\phi_n \sin\phi_{n+k}) \Big] \nonumber \\
- & 2K_1 \cos\theta_n \sin\phi_n \cos\phi_n-\mu B^y \cos\phi_n.
\end{align}
In this last equation a global factor of $\cos\theta_n$ canceled out. These equations give the general lattice dynamics.  
Due to their complexity, a solution of the time-dependence likely requires numerics.  However, they recover the 
three static solutions found previously ($x$-parallel/oblique, $y$-parallel, and $y$-alternating).

\subsection{Verification of static uniform solutions}
It is apparent that $\dot\phi_n = 0$ results if all $\theta_n=0$. Therefore in a static configuration, 
the dipoles remain pointing in the $xy$-plane.  Then for uniform $\phi_n=\phi$, a constant angle,
the second equation of motion with $\dot\theta_n=0$ becomes the {\em equilibrium equation},
\be
\Big[ \big( 6D \sum_{k=1}^R \frac{1}{k^3} -2K_1 \big) \sin\phi -\mu B^y \Big] \cos\phi = 0.
\ee
This has two solutions.  The first is an {\bf oblique state} where the factor inside the braces is zero:
\be
\label{oblq1}
\sin\phi = \frac{\mu B^y/2}{3\zeta_R D -K_1}, \quad \zeta_R \equiv \sum_{k=1}^R \frac{1}{k^3}.
\ee
The dipoles are near the $x$-direction but tilted towards $y$ by angle $\phi$, due to the field.
The oblique states are doubly degenerate because $\phi$ takes two nonequivalent values.
The other solution type is a {\bf $y$-parallel state}, with
\be
\cos \phi =0, \quad \phi = \pm \pi/2, 
\ee
where all the dipoles point parallel/antiparallel to the $B^y$-field.  
The two $y$-parallel states have different energies in a field.  

Another quasi-uniform state can be found, assuming an alternating structure for the in-plane angles,
\be
\label{yalt}
\phi_n = \pm(-1)^n \frac{\pi}{2}.
\ee
The dipoles point either towards or against the field. Suppose a central site at $n$ has $\phi_n \ge 0$, and its 
neighbors have $\phi_{n \pm 1}=\pi-\phi_n$.  Both sets tilt towards $B^y$. Still using all $\theta_n=0$,
the dynamic equation for $\dot\theta_n$ gives
\be
\Big[ \big( -2D \sum_{k=1}^R \frac{1}{k^3} -2K_1\big) \sin\phi_n -\mu B^y \Big] \cos\phi_n =0.
\ee
Then one solution is indeed the {\bf $y$-alternating state}, with angles as in Eq.\ (\ref{yalt}).
There is another possibility, when the term inside brackets is zero, which gives
\be
\sin\phi_n = -\frac{\mu B^y/2}{\zeta_R D+K_1}.
\ee
This has a smaller angle than the oblique state, and it tilts {\em away from } the field!  
Consideration shows that it is an unstable maximum energy state, of limited importance. 
Therefore, the dynamic angular equations are able to reproduce the three types of uniform states. 

\section{A continuum limit?}
Before considering numerical simulations, a continuum limit will be helpful for the interpretation of
the DWs found there.  Consider solutions assuming smooth behavior, in a continuum limit theory, 
where site $n$ is replaced by position variable $x=na$ and $\phi_n,\theta_n$ are replaced by $\phi(x), \theta(x)$. 
Continuum analysis is needed for the dipolar parts.  To that end, we make approximate expansions 
up to quadratic order in space derivatives around a central site, for instance,
\be
\phi_{n+k} \rightarrow \phi(x+ka) \approx \phi(x)+ka \phi_x+\tfrac{1}{2}(ka)^2 \phi_{xx}.
\ee
where $\phi_x$ and $\phi_{xx}$ are the first and second partial space derivatives. 
This applies to oblique and $y$-parallel states, and not to $y$-alternating.
Do this without linearization in $\phi$, but only in its derivatives.  For example, at a $k$th neighbor,
\begin{align}
& \sin\phi_{n+k} \rightarrow \sin\left[\phi+(ka) \phi_x+\tfrac{1}{2}(ka)^2 \phi_{xx}\right] \\
& \approx \sin\phi \left[ 1-\tfrac{1}{2}(ka)^2 \phi_x^2\right]
+\cos\phi \left[(ka)\phi_x+\tfrac{1}{2}(ka)^2 \phi_{xx}\right]. \nonumber
\end{align}
Terms cubic and higher in derivatives have been dropped. 
Primarily this approach can work only if a few neighbors are included; the continuum limit
is incompatible with accounting for all LRD interactions.  

This can be applied to a pair of terms in $\theta_{n\pm k}$ from the $\dot\phi_n$ equation, where the first
derivatives cancel out
\begin{align}
\sin\theta_{n-k}&+\sin\theta_{n+k}  \nonumber \\
& \approx 2\left[1-\tfrac{1}{2}(ka)^2\theta_x^2\right]\sin\theta+(ka)^2\theta_{xx}\cos\theta.
\end{align}
Proceeding this way, the time derivatives of $\phi(x)$ and $\theta(x)$ can be expressed in
times of their space derivatives.

\subsection{Dynamics of in-plane angle $\phi$}
With $\phi_n\rightarrow \phi(x)$ and $\theta_n\rightarrow \theta(x)$, the continuum dynamic equation for the 
in-plane angle becomes
\begin{align}
\dot{\phi} & = \sum_{k=1}^R \frac{D}{k^3} 
\Big\{ 2\left[1-\tfrac{1}{2}(ka)^2\theta_x^2\right]\sin\theta+(ka)^2\theta_{xx}\cos\theta \nonumber \\
& -\tan\theta \Big[  (-2\cos^2\phi+\sin^2\phi)  \nonumber \\
& \times \left\{ \left[2-(ka)^2(\phi_x^2+\theta_x^2)\right]\cos\theta-(ka)^2 \theta_{xx}\sin\theta \right\}
\nonumber \\
& +3(\sin\phi\cos\phi)(ka)^2 \left[ \phi_{xx}\cos\theta -2\phi_x \theta_x \sin\theta \right] \Big]  \Big\}
\nonumber \\
& +2K_1 \sin\theta \sin^2\phi +2 K_3 \sin\theta +\mu B^y \tan\theta \sin\phi .
\end{align}
The main caveat here is that the sum of $\frac{1}{k}$ is not convergent if $R\to\infty$. Yet we expect that a real
system would have a finite and well-defined solution. Taking account of LRD interactions this way into
a continuum limit is clearly risky.  Probably this is a model that will be useful up to a few nearest neighbors. 
One can see, however, that a choice $\theta(x)=0$ completely makes the RHS vanish, giving a static solution, $\dot\phi=0$,
independent of the actual $x$-dependence of $\phi$.

\subsection{Dynamics of out-of-plane angle $\theta$}
Following a similar procedure, the resulting dynamic equation for $\dot\theta$ becomes 
\begin{align}
\label{dot-theta}
\dot\theta(x) & = \sum_{k=1}^R \frac{D}{k^3} \Big\{ \nonumber \\
& (-2\sin^2\phi+\cos^2\phi)(ka)^2 \{ \phi_{xx}\cos\theta -2\phi_x \theta_x \sin\theta \} \nonumber \\
& +(3\sin\phi\cos\phi) \nonumber \\
& \times \left[ \left\{ 2-(ka)^2(\phi_x^2+\theta_x^2) \right\} \cos\theta -(ka)^2 \theta_{xx} \sin\theta \right] \Big\}
\nonumber \\
& -2K_1 \cos\theta \sin\phi\cos\phi -\mu B^y \cos\phi.
\end{align}
Again, the summations of $\frac{1}{k}$ over long-range are not convergent, and this will only make sense with a few
nearest neighbors. Initially, I will consider the static solutions that are possible for the NN limit.  
Then, longer range interactions will modify that.  A starting point is assuming $\theta=0$, which implies 
$\dot\phi=0$, and then use that in conjunction with $\dot\theta=0$, which gives a nonlinear equilibrium equation 
to be solved under given boundary conditions. 

\begin{figure}
\includegraphics[width=\figwidth,angle=0]{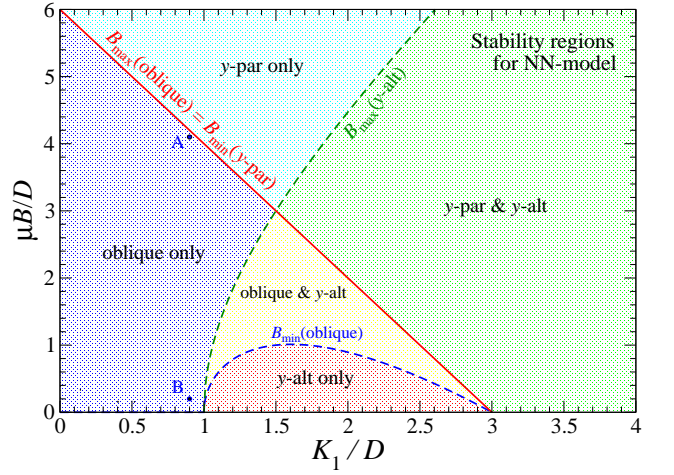}
\caption{\label{nn-phases} Regions in the anisotropy-applied field coordinates $(k_1,b)$ where the different
uniform states are stable in the NN model. In relaxation simulations, point A produced the smooth oblique--oblique 
DW in Fig.\ \ref{oblq-nn-k090-b41}; point B produced the oblique--oblique DW with AFM order in Fig.\ 
\ref{obliq-nn-af-k090-b02}. }
\end{figure}

\subsection{Static NN limit}
The simplest problem is to seek the static soliton-like solutions in the NN model, connecting oblique states 
and maybe also $y$-par states. Connections with $y$-alt states requires a two-sublattice calculation, not
considered here. 

If one wants $\dot{\phi}(x)=0$, it can be seen in Eq.\ (\ref{dotphi}) that even without a continuum
approximation or linearization, $\theta\equiv 0$ is sufficient:
\be
\theta(x) = 0 \quad \implies \quad \dot{\phi}(x) = 0. \quad \text{Static solutions.}
\ee
Then in a static situation, with $\theta=\dot{\theta}=0$, Eq.\ (\ref{dot-theta}) gives a nonlinear differential
equation for equilibrium configurations. Using the near-neighbor approximation to the dipole interactions
(range $R=1$), it is
%
\begin{align}
\label{nn0}
(1- & 3\sin^2\!\phi)\phi_{xx} \nonumber \\
& +\left[(6-2k_1-3\phi_{x}^2) \sin\phi - b\right] \cos\phi = 0.
\end{align}
For simplicity $x$ is now measured in units of the island spacing, $a$.  That means $x$ is equivalent 
to island index $n$, but I use $x$ to indicate the continuum theory.  I also use reduced 
anisotropy and field constants, relative to the dipole coupling strength, defined as
\be
k_1 \equiv \frac{K_1}{D}, \quad b \equiv \frac{\mu B^y}{D}.
\ee
One can verify that the equation has two uniform solutions, setting $\phi_x=\phi_{xx}=0$,
\be
\label{sinphi}
\sin\phi = \frac{b/2}{3-k_1} \quad \text{ or } \quad \cos\phi = 0,
\ee
which are the oblique solutions with just NN interactions [compare Eq.\ (\ref{oblq1})
with LRD interactions] and the $y$-parallel solutions.  Enforcing the constraint, 
$|\sin\phi|\le 1$, in the NN model the maximum field strength for which there are
oblique states is 
\be
\label{bmaxnn}
b_{\rm max} = 2(3-k_1).
\ee
We see below that $b$ relative to $b_{\rm max}$ is relevant for determining DW widths.  
Fig.\ \ref{nn-phases} shows the stability regions of the three types of uniform states when this model is limited to
NN interactions, according to the linear stability analysis in Ref. \cite{Wysin24}. Eq.\ (\ref{bmaxnn}) is depicted as 
a red line in Fig.\  
\ref{nn-phases}.

\subsection{Static continuum energy in NN model}
The energy required to create a domain-wall is certainly of major importance. Towards that end, 
the continuum limit can be applied to the NN Hamiltonian using the same procedures as those for the dynamic equations.  
Putting $\theta=0$ throughout the system (dipoles remain pointing in the $xy$-plane), the static NN Hamiltonian
can be expressed as
\begin{align}
\label{Hnn-phi}
H = D \int dx \big\{ & (1-3\cos^2\phi)+\left(1-\tfrac{3}{2} \cos^2\phi \right) \phi_{x}^2 \nonumber \\
&  -k_1\sin^2\phi-b\sin\phi \big\}.
\end{align}
An assumption $\phi_x\to 0$ at $x=\pm \infty$ was used with integration by parts to convert a
term depending on $\phi_{xx}$ into one with $\phi_x^2$.  One can verify that the Euler-Lagrange
equation for minimizing this energy is indeed the equilibrium equation (\ref{nn0}). Later this
is used to estimate the DW energy.

\section{Oblique-to-oblique domain walls in the NN model}
The oblique states always have a two-fold degeneracy, corresponding to two opposite values for $S^x=\cos\phi$.
From Eqs.\ (\ref{sinphi}) and (\ref{bmaxnn}), the longitudinal spin component in a uniform oblique state can be
\be
S^x = \cos\phi_{\rm oblq}^{\pm} = \pm\sqrt{1-\left(\frac{b}{b_{\rm max}}\right)^2}.
\ee
A domain wall will connect from the $S^x>0$ state to the $S^x<0$ state. The change in the
angle $\phi$ will be greatest ($180^{\circ}$) for the case of zero field, $b=0$.  At the
opposite extreme, if the field is near the maximum allowed for an oblique state, the change
in $\phi$ will be very small, and perhaps the domain wall is hardly noticeable. For example,
changing from $\phi_{\rm oblq}^{+}=85^{\circ}$ with $S^x>0$ to $\phi_{\rm oblq}^{-}=95^{\circ}$ 
with $S^x<0$.  If $b=0$, the dipoles may also rotate either clockwise or counterclockwise while 
moving down the chain towards increasing $x$.  But if $b>0$, one expects that only a rotation where 
the dipoles tilt towards the field direction will tend to be stable.

\subsection{Analytic solution for oblique--oblique DWs, NN model}
\begin{figure}
\includegraphics[width=\figwidth,angle=0]{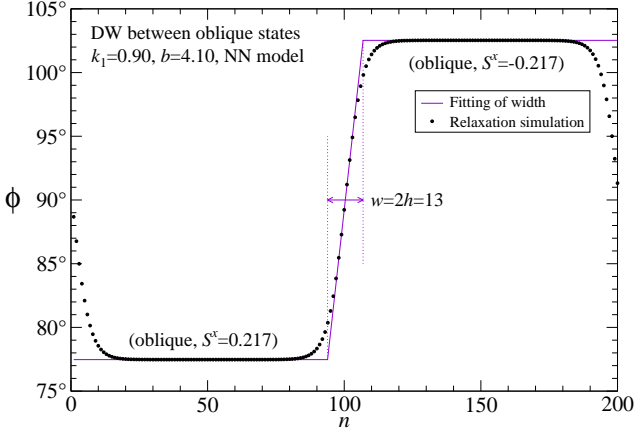}
\caption{\label{oblq-nn-k090-b41} For the NN model with parameters at point A in Fig.\ \ref{nn-phases}, 
a DW between oblique states and the fitting of the tangent line at its center.  
This is a case where the maximum magnetic field for oblique states is $b_{\rm max}=4.20$ .
The width is $w=2h$ where $h=6.5$ is the half-width, measured in units of the island spacing, $a$.  
The $\varphi^4$ theory, Eq.\ (\ref{hbeta}), gives $h=6.32$ for this situation.}
\end{figure}

Via numerical relaxation simulations, below, I noticed that oblique--oblique DWs do exist and 
have a soliton shape $\phi(x)$ reminiscent of those with a hyperbolic tangent dependence on $x$,  
such as that in Fig.\ \ref{oblq-nn-k090-b41}.
The function $\phi(x)$ is symmetric around the point $\phi=\pi/2$, with $\phi$ being either above or below
$\pi/2$ by a small amount when the applied field $b$ is near its maximum $b_{\rm max}$ for stable oblique 
states.  Further, the width of the obtained DWs grows larger as $b$ approaches $b_{\rm max}$, and
the dipole configuration becomes much better approximated as a continuous function.  
This led me to consider this regime as a limiting case and solve the problem with a new variable, $\psi$, 
that is better for expansion.  The new variable must be the deviation of $\phi$ from $\pi/2$, i.e., define
\be
\phi=\frac{\pi}{2}+\psi \quad \implies \quad \psi\equiv \phi-\frac{\pi}{2}.
\ee
The oblique state with $S^x>0$ has $\psi<0$ and the oblique state with $S^x<0$ has $\psi>0$.
The NN continuum equation (\ref{nn0}) is not linearized in $\phi$, but only assumes smooth derivatives. 
Using $\sin\phi = \cos\psi$ and $\cos\phi=-\sin\psi$, it can be converted to the equivalent equilibrium 
equation for $\psi$,
\be
\label{psi-eq}
(-2+3\sin^2\psi)\psi_{xx}-(b_{\rm max}-3\psi_x^2)\cos\psi \sin\psi +b\sin\psi = 0. 
\ee
One can check that it reproduces the uniform oblique states, using $\psi_x=\psi_{xx}=0$,
which gives
\be
\sin\phi = \cos\psi = \frac{b}{b_{\rm max}}.
\ee
If one does consider $\psi \ll 1$, this leads to its approximate values in the oblique states,
\be
\label{psiX}
1-\tfrac{1}{2}\psi^2 \approx \tfrac{b}{b_{\rm max}} \quad \implies \quad
\psi \approx \pm \sqrt{\tfrac{2(b_{\rm max}-b)}{b_{\rm max}}}.
\ee
This clearly shows that $\psi$ becomes small in the limit $b \to b_{\rm max}$, suggesting
that expansion in $\psi$ can be of value also for DW solutions.  

\subsubsection{Expansion into a $\varphi^4$ problem}
\label{ssphi4}
For $b$ close enough to $b_{\rm max}$, expansion in $\psi\ll 1$ should be reliable. I will start from
the NN continuum energy density $u$ in units of $D$, obtained from (\ref{Hnn-phi}) and expressed in
terms of $\psi$:
\begin{align}
\label{upsi}
u  = & -2+\left(1-\tfrac{3}{2} \sin^2\psi \right) \psi_{x}^2 \nonumber \\
 & +[(3-k_1)\cos\psi-b]\cos\psi .
\end{align}
Now I keep terms up to quartic order of smallness, where $\psi$ is first order, $\psi_x$ is second order, and
so on.    That means that $(\sin^2\!\psi)\psi_{x}^2$ is already sixth order and can be dropped. Using 
up to quadratic terms in the expansions of cosine, we have
\be
u  \approx -2+\psi_{x}^2 +\left[(3-k_1)(1-\tfrac{1}{2}\psi^2)-b \right](1-\tfrac{1}{2}\psi^2) .
\ee
Collecting terms, we can write this compactly as
\be
\label{u4}
u  \approx u_{y\text{-par}}+\psi_{x}^2 -A\psi^2 +\tfrac{1}{2}B\psi^4,
\ee 
where $u_{y\text{-par}}$ is the energy density in a $y$-parallel state ($\psi=\psi_x=0$, 
dipoles aligned with $B^y$),
\be
u_{y\text{-par}} = -2+\tfrac{1}{2}b_{\rm max}-b
\ee
and the constants are
\be
A \equiv \tfrac{1}{2}\left(b_{\rm max}-b\right), \quad B\equiv \tfrac{1}{4}b_{\rm max}.
\ee
In Eq.\ (\ref{u4}), $u$ is the energy density of a $\varphi^4$ model (for variable $\psi$), a classical 
scalar field theory containing an elastic-like energy ($\psi_x^2$) combined with a double-well potential, 
see Ref.\ \cite{Phi4Notes} for an introductory description. 
This case does not contain a kinetic energy term like $\psi_{\tau}^2$ because only the static
solutions are considered.  The two minima of the potential at $\psi=\pm \sqrt{A/B}$ represent 
the two oblique states with opposite $x$-components of magnetization, compare Eq.\ (\ref{psiX}). 
The constant, $u_{y\text{-par}}$, is the background energy density per site for all the dipoles 
pointing in the $y$-direction.

\subsubsection{Cubic order equilibrium equation and its kinks}
The $\varphi^4$ equilibrium equation can be obtained from Eq.\ (\ref{psi-eq}) consistently
with the energy density in (\ref{u4}) by using the quadratic expansion of $\cos\psi$
together with the linear expansion of $\sin\psi$, limiting to cubic terms, resulting in
%
\be
-2\psi_{xx}-b_{\rm max}(1-\tfrac{1}{2}\psi^2)\psi +b\psi \approx 0.
\ee 
This can be expressed as
\be
\label{phi4}
-2\psi_{xx}+(b-b_{\rm max})\psi +\tfrac{1}{2}b_{\rm max}\psi^3 \approx 0.
\ee
I will refer to Eq.\ (\ref{phi4}) as the $\varphi^4$ equilibrium equation and $\varphi^4$ theory for this magnetic problem.
It can be verified from the Euler-Lagrange minimization of the energy density $u$ in (\ref{u4}), or its 
functional derivative,
\be
\frac{\partial u}{\partial \psi}-\frac{d}{dx} \left( \frac{\partial u}{\partial \psi_x}\right) = 0,
\ee
which produces a simple form in terms of the previously defined constants $A$ and $B$,
\be
\label{psiAB}
-\psi_{xx}+A\psi+B\psi^3 =0.
\ee
This is the simplest consistent and nontrivial expansion.  The next higher approximation (up to 
quartic order in $\psi$ for $\cos\psi$ in $u$) would produce up to seventh-order terms in the
equilibrium equation, which will not be tractable.

Eq.\ (\ref{psiAB}) has static soliton solutions, commonly referred to as kinks, that represent domain-walls. 
When centered at $x=0$, a static kink is
\be
\label{tanhx}
\psi = C \tanh \beta x, 
\ee
where amplitude $C$ and inverse length $\beta$ are to be determined.  Taking its space derivatives, one has
\begin{align}
\psi_x & = C \beta \sech^2 \beta x, \nonumber \\
\psi_{xx} &= -2 C\beta^2 \sech^2 \beta x\, \tanh \beta x.
\end{align}
Using these in Eqs.\ (\ref{psiAB}) leads to two constraints that determine the constants,
\begin{align}
BC^2 &= A 
\quad \implies \quad C=\sqrt{\tfrac{A}{B}} = \sqrt{\tfrac{2(b_{\rm max}-b)}{b_{\rm max}}}, \\
2\beta^2 &= A 
\quad \implies \quad \beta = \sqrt{\tfrac{A}{2}} = \tfrac{1}{2}\sqrt{b_{\rm max}-b}.
\end{align}
%
$C$ is indeed the amplitude needed to match to the uniform oblique states at 
$x= \pm\infty$, Eq.\ (\ref{psiX}).   $\beta$ determines the DW width.

\subsubsection{DW creation energy, NN $\varphi^4$ model}
Based on the continuum Hamiltonian (\ref{Hnn-phi}) approximated with the $\varphi^4$ energy density in 
Eq.\ (\ref{u4}), the energy needed to create a DW between two equal-energy oblique states can be estimated. 
Using the kink solution in the $\varphi^4$ energy density of Eq.\ (\ref{u4}), it becomes 
\be
u = u_{y\text{-par}}-\tfrac{1}{2}AC^2+ AC^2 \sech^4 \beta x,
\ee 
where 
\be
AC^2 = \frac{\left(b_{\rm max}-b\right)^2}{b_{\rm max}}.
\ee
The net background constant is found to be the energy density of an oblique state,
\be
u_{y\text{-par}}-\tfrac{1}{2}AC^2 = u_{\rm oblq} = -2-\frac{b^2}{2 b_{\rm max}}.
\ee 
Then, the energy {\em above} the oblique background is the creation energy of a DW between
oblique states. Stated otherwise, the creation energy is the difference in energy of the
system with a DW present, minus the energy of the system with a uniform oblique state present. 
For the DW creation energy, the integration of $\sech^4\beta x$ over all $x$ produces 
\be
\int_{-\infty}^{\infty} \sech^4\beta x \ dx = \frac{4}{3\beta}.
\ee
Then the resulting DW creation energy in units of $D$ is obtained only from this part of $u$, 
\begin{align}
\label{EDW}
\frac{E_{\rm DW}}{D} & = AC^2 \int_{-\infty}^{\infty} \sech^4\beta x \ dx \nonumber \\
& = \frac{4\sqrt{2}}{3}\frac{A^{3/2}}{B} = 
\frac{8}{3} \frac{\left(b_{\rm max}-b\right)^{3/2}}{b_{\rm max}}.
\end{align}
This correctly predicts a vanishing creation energy when $b$ reaches $b_{\rm max}$, where oblique becomes
a $y$-par state.  The expected $3/2$ power law will be checked below for the DWs obtained from relaxation 
simulations.

\subsection{Analytic and geometric definitions of the DW width}
The simplest definition for width is just the inverse of $\beta$, but it will be good to relate it
to the graphs of numerical results.
I did a fitting of a width parameter by finding the intersection of the tangent line at 
the DW center with the constant line at the asymptotic value of $\phi$ or equivalently, $\psi$. 
See examples of the tangent line and constant lines in Fig.\ \ref{oblq-nn-k090-b41}.
The slope of the central tangent line is
\be
\psi_x(0) = \phi_x(0) = C\beta.
\ee
The tangent line through the DW center is defined by 
\be
y(x) = \psi_x(0) x = C\beta x.
\ee
It intersects the constant asymptotic value, $\psi(\infty) = C$, at a half-width $h$.  That gives 
easily and not surprisingly,
\be
y(h) = C\beta h = C \quad \implies \quad h = \text{half-width} = \beta^{-1}.
\ee 
The half-width obtained from the tangent line intersection is an estimate of the parameter $\beta^{-1}$,
I also define the full width as $w=2h$.  Then the  half-width fitted from simulations can be compared with the 
analytic prediction from the $\varphi^4$ equation,
\be
\label{hbeta}
h = \beta^{-1} = \frac{2}{\sqrt{b_{\rm max}-b}}.
\ee
It predicts a divergence of the DW width as one moves from the oblique states into the
region where only $y$-parallel states are possible.  

\begin{figure}
\includegraphics[width=\figwidth,angle=0]{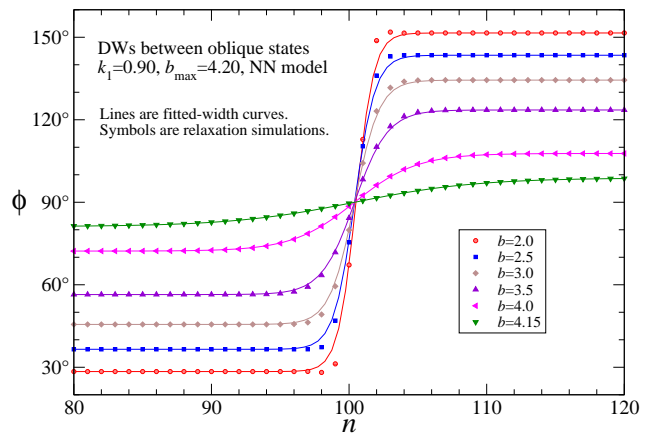}
\caption{\label{oblq-nn-k090-multi} For the NN model, DWs between oblique states for a low value of anisotropy, 
$k_1=0.90$, for a range of transverse applied field $b$. Points are the relaxation simulation; curves are the
$\varphi^4$ theory using fitted widths.  The system size is $N=200$, and the view is zoomed into the
center where the DW is.  There is an obvious trend in the DW width as $b$ approaches the maximum, where
the system will go into a $y$-par state.}
\end{figure}

\begin{figure}
\includegraphics[width=\figwidth,angle=0]{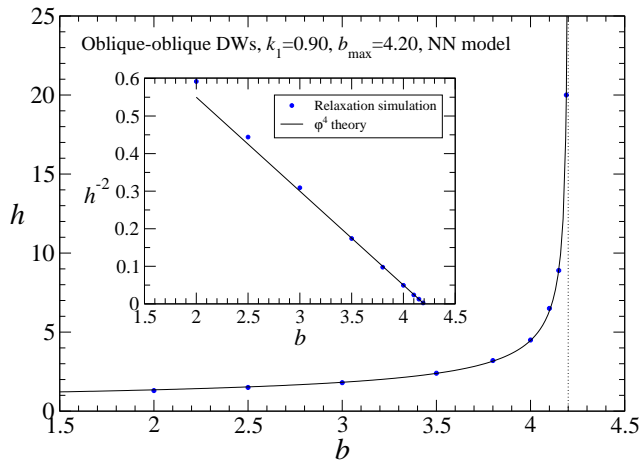}
\caption{\label{h-obliq-k090+phi4} For the NN model, DW half-widths between oblique states for a low value of anisotropy, 
$k_1=0.90$, as a function of transverse applied field $b$.  Points are relaxation simulation data while the curves are
the $\varphi^4$ theory, Eq.\ (\ref{hbeta}). The insert shows a linear dependence of $\beta^2=h^{-2}$ on $b$
until $b$ reaches $b_{\rm max}$.}
\end{figure}

\subsection{Relaxation simulations of oblique-oblique DWs, NN model}
Relaxation simulations as described in Sec.\ \ref{sims} were performed for the NN model, putting $R=1$ 
in the dipole sums, for tests of the $\varphi^4$ theory. 
I used systems mostly with $N=200$ islands, except for cases with $b$ close to $b_{\rm max}$ where the 
DW width required a larger systems ($N=500$).  Oblique-oblique DWs are reliably reproducible if anisotropy 
$k_1$ is small and the applied field is approaching its maximum value for oblique states.  The easy-plane 
anisotropy parameter $k_3$ was set to zero. Any positive value of $k_1$ already imposes its own effective 
easy-plane tendency that results in $\theta=0$ for the static configurations.  

\subsubsection{DW configurations}
An example was shown in Fig.\ \ref{oblq-nn-k090-b41} for $k_1=0.90$, where half the system was started 
in one oblique state and half the system was in the other oblique state.   The plot shows the result 
after relaxation over several thousand iterations, for an open-ended system.  A DW forms to match the 
two regions, as expected.  There is also a half-DW at each end of the open-ended chain, apparently due to 
the missing dipole interactions at the ends.  The example shown has $b$ very close to $b_{\rm max}$, 
which gives a significant width $w=2h$=13, as found from the fitting lines.  This type of result 
appears for $b\ge 2.0$ .  For weaker $b$ more complex nonuniform configurations result that exhibit
alternating dipoles, see below for examples.

For $k_1=0.90$, the trends in the DW for varying $b$ are shown in Fig. \ref{oblq-nn-k090-multi}.
It becomes obvious that the width increases as $b$ approaches close to its maximum value, while
the amplitude of the structure reduces.  It is clear that the continuum approach
becomes more reliable as $b\to b_{\rm max}$.  The fitted $\varphi^4$ theory curves (using the fitted widths $h$) 
follow the relaxation data very well for the higher values of $b$.  For smaller field $b$, the fitted 
$\varphi^4$ theory curves deviate from the relaxation simulation data at the shoulders of the DW. 
In those cases, however, the widths are very small ($h \sim 1$) and one cannot expect the continuum 
theory to hold well. 

\begin{figure}
\includegraphics[width=\figwidth,angle=0]{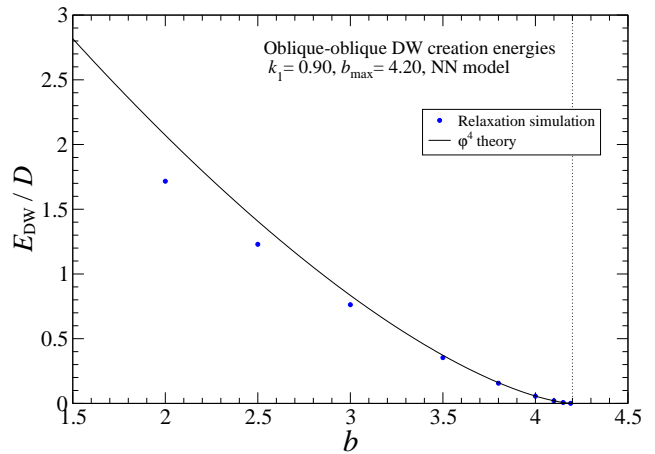}
\caption{\label{E-nn-k090} For the NN model, DW creation energy between oblique states for a low value of anisotropy,
$k_1=0.90$, as a function of transverse applied field $b$.  Points are relaxation simulation data while the curves are
the $\varphi^4$ theory, Eq.\ (\ref{EDW}). The simulated creation energy is the difference in energies of the system with a DW
present minus the system in a relaxed oblique state.}
\end{figure}

\subsubsection{DW half-widths and creation energies}
The results for the fitted half-widths are shown in Fig.\ \ref{h-obliq-k090+phi4}, where points are
the relaxation simulation data and the curves are from the $\varphi^4$ theoretical expression, Eq.\ (\ref{hbeta}),
without any fitting parameters.  The data follow the theoretical curve extremely well.  The insert shows a
different view of the relationship.  These data confirm that the approximations used to get the 
$\varphi^4$ theory are valid and reliable.  The changing structures also indicate what can happen to the
oblique states with increasing applied field, as the system will transition into a global $y$-par state.

The $\varphi^4$ theory creation energy is compared with the relaxation results in Fig.\ \ref{E-nn-k090}.
The theory works very well for $b$ close to $b_{\rm max}$.  Further out, the theoretical DW energy is somewhat
higher than the simulations.  It implies that the relaxation is able to achieve a slight deformation away
from the approximate $\varphi^4$ shape,  and that advantageous deformation lowers the energy. It confirms that
the $\varphi^4$ theory is good for describing the DWs when there are only NN interactions.

In a later section, the NN continuum approximation is modified somewhat to roughly describe the 
oblique-oblique DWs for the LRD model.  Next, another type of DW is described for the NN model.

\subsection{Oblique--oblique DWs with antiferromagnetic order in the NN model}
One can note the reason why $b<2.0$ was not used above for the oblique-oblique DWs.  If the field is
too small, antiferromagnetic (AFM) ordering creeps in, especially at the chain ends.  But it
can also affect the DW structure, that then does not stay as a smooth $\tanh\beta x$ form. Rather, 
there are two smooth curves on each sublattice of even or odd sites.  It is hinted at in the shoulders 
of the DWs in Fig.\ \ref{oblq-nn-k090-multi} for small $b$. This is because NN dipole interactions tend 
to make neighboring dipoles acquire AFM order. 

\begin{figure}
\includegraphics[width=\figwidth,angle=0]{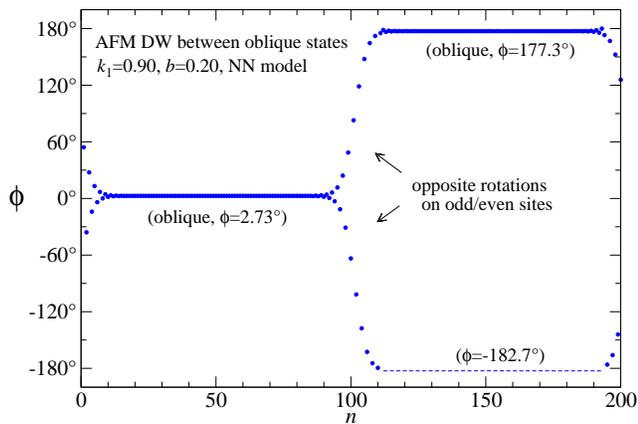}
\caption{\label{obliq-nn-af-k090-b02} Example from relaxation simulations with a weak applied field 
that can hold a DW with AFM order between two oblique states (parameters at point B in Fig.\ \ref{nn-phases}). 
Note the dashed line at $\phi=-182.7^{\circ}$ (physically the same as $\phi=177.3^{\circ}$) suggests the continuation 
of the oblique state at odd sites. The DW creation energy is $E_{\rm DW} = 1.176 D$.} 
\end{figure}

An interesting example of the AFM ordering is shown in Fig.\ \ref{obliq-nn-af-k090-b02} for
$k_1=0.90$ and a very weak but nonzero field.  Oblique states are the only stable uniform states for
these parameters.  Here the two oblique states are linked via a DW itself with AFM order!  
Also, the even and odd site dipoles rotate in opposite directions in going from one oblique state
to the other.  Such a nonuniform state can be metastable, and likely owes its stability to the dipole 
interactions.  The NN dipolar effects drive the AFM ordering of neighboring islands, and the applied 
field is too weak to counteract that phenomenon. The DW creation energy in this configuration is 
$E_{\rm DW} = 1.176 D$, which is small compared to the $\varphi^4$ prediction from Eq.\ (\ref{EDW}),
$E_{\rm DW} = 5.079 D$, if AFM order were not present in the DW.

\begin{figure}
\includegraphics[width=\figwidth,angle=0]{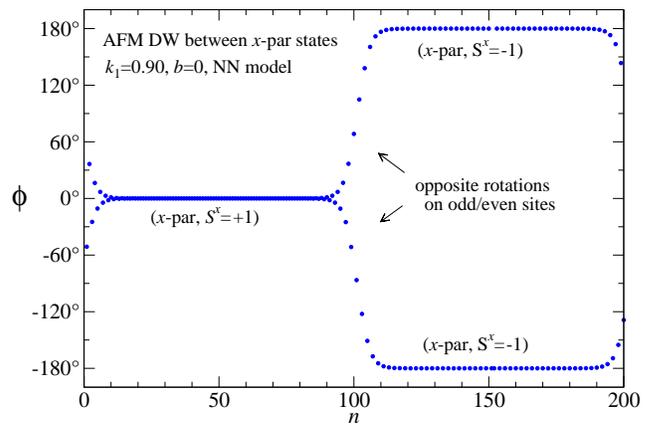}
\caption{\label{obliq-nn-af-k090-b0} Example from relaxation simulations with $b=0$ applied field 
that can hold a DW with AFM order between the two $x$-par states. Note that $\phi=+180^{\circ}$
and $\phi=-180^{\circ}$ are physically equivalent $x$-par states.  The DW creation energy is 
$E_{\rm DW} = 1.145 D$.}
\end{figure}

Another example is shown for $b=0$ in Fig.\ \ref{obliq-nn-af-k090-b0}.  
There is a DW with AFM order connecting the two oblique states, that are really $x$-parallel
states with $S^x = +1$ on the left side and $S^x= -1$ on the right side.  Note that the angles
$\pm 180^{\circ}$ measured from the $+x$-axis are the same direction, towards $-x$. If there is any
AFM order modulating these oblique states, it is of extremely small amplitude. The DW creation 
energy is $E_{\rm DW} = 1.145 D$, also small compared to the $\varphi^4$ prediction from Eq.\ (\ref{EDW}),
$E_{\rm DW} = 5.465 D$, without AFM order. 

These examples indicate that AFM ordering is a physical exhibition of the NN dipolar interactions.
While AFM order is being established, the 2nd NN dipole interactions (not in these examples) would 
tend to oppose the site-to-site AFM order, but with 1/8 the strength of NN interactions.  Then, 
a system with LRD may behave somewhat differently. Next the DWs in the LRD model are explored.

\section{Relaxation simulations of DWs in the LRD model}
\begin{figure}
\includegraphics[width=\figwidth,angle=0]{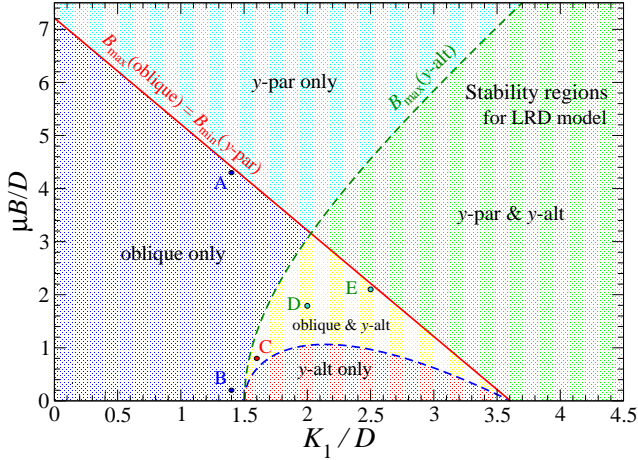}
\caption{\label{lrd-phases} Anisotropy-applied field coordinates $(k_1,b)$ where the uniform states are stable 
in the LRD model, from Ref.\ \cite{Wysin24}. In relaxation simulations, point A produced the smooth oblique--oblique
DW in Fig.\ \ref{oblq-lrd-k14-b43}, point B produced the oblique--oblique DW with AFM order in Fig.\
\ref{oblq-lrd-k14-b02}, point C produced the $y$-alt--$y$-alt DWs in Figs.\ \ref{alt-kink-k16-b08-a} and
\ref{alt-kink-k16-b08-b}, and points D and E correspond to the oblique--$y$-alt DWs in Figs. 
\ref{ob-yalt-lrd-k20-b179} and \ref{ob-yalt-lrd-k25-b21}.}
\end{figure}

Eq.\ (\ref{nn0}) fairly accurately describes the NN model, and further, Eq.\ (\ref{phi4}) even includes 
enough nonlinearity to accurately describe the oblique--oblique DWs. To be realistic, however, the LRD 
interactions must be included.  Following below are numerical results for systems of $N=200$ islands 
while including all the LRD interactions, by choosing the upper limit of the dipole sums as $R=N$ for 
an open-ended system, From linear stability analysis in Ref.\ \cite{Wysin24}, the stability regions of 
the three types of uniform states in the LRD model are shown in the $(k_1,b)$ plane in Fig.\ \ref{lrd-phases}. 

\subsection{Oblique--oblique DWs in the LRD model}
\begin{figure}
\includegraphics[width=\figwidth,angle=0]{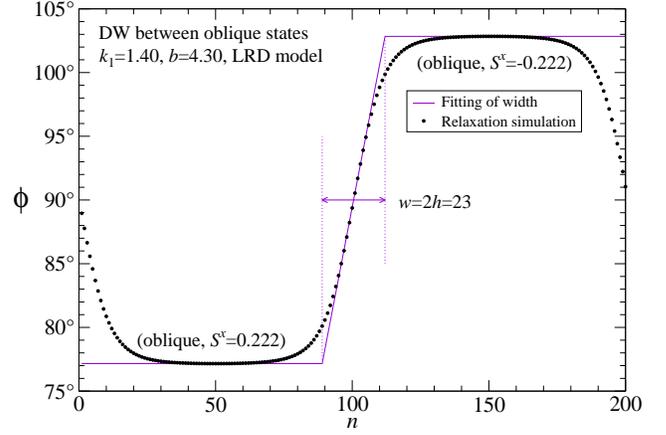}
\caption{\label{oblq-lrd-k14-b43} For the LRD model at point A in Fig.\ \ref{lrd-phases}, a DW between oblique 
states and the fitting of the tangent line at its center.  This is a case where the maximum magnetic field for 
oblique states is $b_{\rm max}=4.412$ .  The width is $w=2h$ where $h=11.5$ is the half-width.  
The $\varphi^4$ theory for the NN model, Eq.\ (\ref{hbeta}), gives $h\approx 6$ for this situation.}
\end{figure}

For oblique DWs with $k_1=1.40$, the results are similar to those for the NN model with $k_1=0.90$,
because it is below but near the maximum $k_1$ for only oblique stable states at zero field.
A general example of a fit for the half-width $h$ in terms of the $\tanh(x/h)$ function, is
given for field $b=4.3$ in Fig.\ \ref{oblq-lrd-k14-b43}.  The maximum field is $b_{\rm max}=4.412$ 
for oblique states at this anisotropy; they go over to $y$-par beyond that. Note that fitting line
matches the slope at the center of the DW. The amplitude is already determined by the angle $\phi$
in the two oblique states, symmetrically displaced about $90^{\circ}$. If the NN $\varphi^4$ theory
of Eq\ (\ref{hbeta}) is tried, it gives for the half-width, 
\be
h= \frac{2}{\sqrt{b_{\rm max}-b}} = \frac{2}{\sqrt{4.412-4.3}}= 5.98
\ee
but the fitted half-width is $h=11.5$, about twice as large.  We can sketch physically why this is so. If
neighboring dipoles are nearly aligned and perpendicular to the chain direction in the DW center, it raises 
the dipole energy, compared to the dipoles in a uniform oblique state. The dipoles in the DW center tilt
towards  the $b$-field; that energy is lowered if the width is larger.  Including LRD interactions raises 
the dipole energy more than with just NN interactions. Then, with LRD interactions, the raised dipole energy can 
be compensated with more (negative) $b$-field energy if the DW becomes wider.  Regardless of the exact process,
the DW creation energy is $E_{\rm DW} = 0.0477 D$, quite small because $b$ is very close to $b_{\rm max}$, 
but larger than the NN $\varphi^4$ theory, $E_{\rm DW} = 0.0226 D$, Eq.\ (\ref{EDW}). 

%
\subsubsection{DW shapes}
\begin{figure}
\includegraphics[width=\figwidth,angle=0]{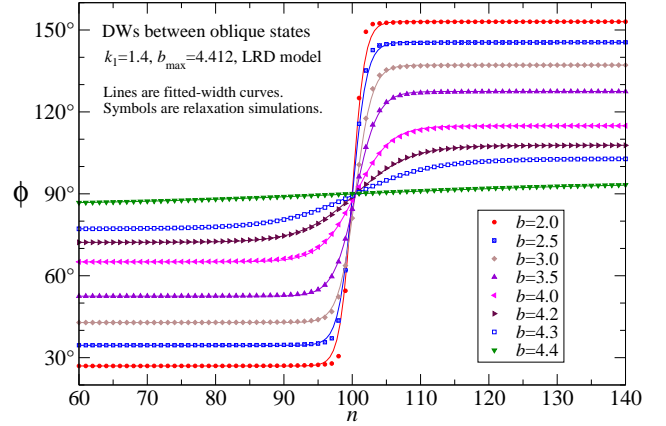}
\caption{\label{oblq-lrd-k14-fits} For the LRD model, DWs between oblique states for a value of anisotropy,
$k_1=1.40$, for a range of transverse applied field $b$. Points are the relaxation simulation; curves are the
$\varphi^4$ expression (\ref{phi4-psi}) using fitted widths.  The view is zoomed into the center where the DW is, for
systems with 200 to 500 islands.  As $b$ approaches the maximum $b_{\rm max}=4.412$, the system will go into 
a $y$-par state.}
\end{figure}

As for the NN model, the shape of the DWs is not exactly of the form $\tanh(x/h)$.  
For $k_1=1.40$, a set of the DWs is shown in Fig.\ \ref{oblq-lrd-k14-fits}, for a range of different field 
$b$. The fitted curves have the needed slope at DW center, but do not follow the DW shoulders very well, 
especially for the narrowest DWs.  Perhaps that is simply a breakdown of the continuum limit there. 
Otherwise, the theory actually works well, although it needs a prediction for the half-width. 

\subsubsection{Weak field}
\begin{figure}
\includegraphics[width=\figwidth,angle=0]{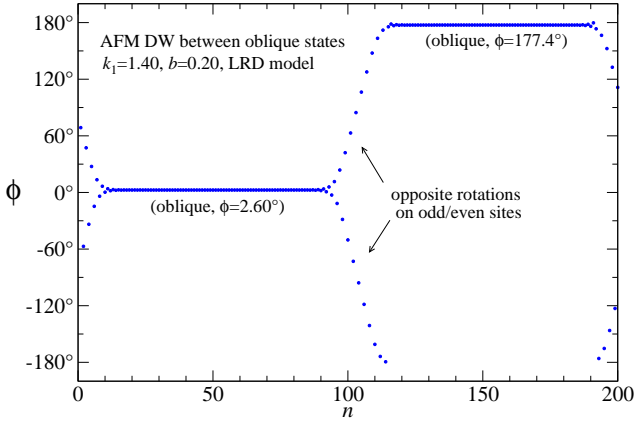}
\caption{\label{oblq-lrd-k14-b02} For the LRD model with parameters at point B in Fig.\ \ref{lrd-phases}, 
a DW with AFM local order connects the two oblique states. The creation energy is $E_{\rm DW} = 2.081 D$. }
\end{figure}

A similar test is to use a field quite less than $b=2$, where AFM tendencies are apparent.
An example is shown in Fig.\ \ref{oblq-lrd-k14-b02} for $k_1=1.40$ and $b=0.20$, including LRD interactions.
The DW with AFM order seems to be very stable, even though the $(k_1,b)$ parameters are outside the stability 
range of a uniform AFM state such as $y$-alt.  Of course, it is important to point out that the alternation
inside the DW {\em is not} a $y$-alt state, clearly since it is not uniform nor with angles at $\pm 90^{\circ}$.
The creation energy is $E_{\rm DW} = 2.081 D$.

\subsubsection{Zero field}
\begin{figure}
\includegraphics[width=\figwidth,angle=0]{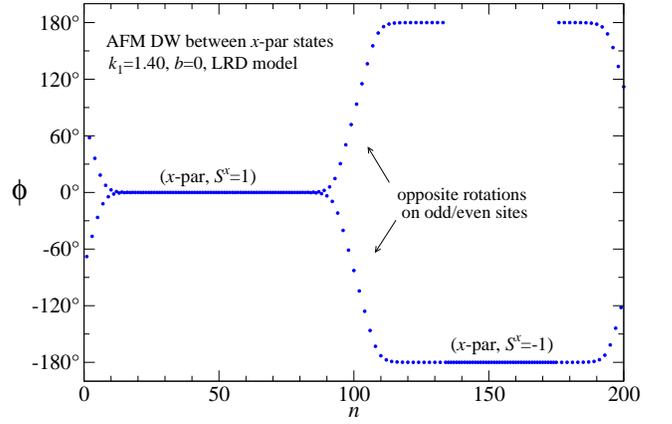}
\caption{\label{oblq-lrd-k14-b00} For the LRD model with $b=0$, a DW with AFM local order connects the
two $x$-par states (i.e., oblique states with $\phi=0, 180^{\circ}$) for $k_1=1.40$. Note that
$\phi=-180^{\circ}$ and $\phi=+180^{\circ}$ are physically equivalent to the same $x$-par state.
The creation energy is  $E_{\rm DW} = 2.038 D$.}
\end{figure}

With the field turned off ($b=0$), there can likely be some different outcomes. This is
because of the tendency for NN dipole interactions to cause the dipoles to go to an AFM configuration. As
seen earlier, this occurs more easily inside a DW, or near the ends of the island chain.  An example
including LRD interactions is shown in Fig.\ \ref{oblq-lrd-k14-b00}. On the left is one oblique state ($x$-par
with $\phi=0$), connected through an AFM-ordered DW region to the opposite oblique state
($x$-par with $\phi=-180^{\circ}$, physically equivalent to $\phi=+180^{\circ}$).  I did many simulations
with different initial states. As long as there are different starting angles in the two sides of
the system, it relaxes into this configuration, 
with a DW creation energy $E_{\rm DW} = 2.038 D$.

\subsubsection{Half-widths} 
We can summarize the DW half-widths while including LRD interactions for $k_1=1.40$ .
The simulation data are shown as points in Fig.\ \ref{h-obliq-k14+phi4}.  There is a fast increase 
in width as $b$ approaches its maximum value for oblique DWs, similar to the curve for the NN model.  
However, the general sizes of the widths are larger here.  The inverse square half-width has close to 
a linear relation with $b$, however, it is slightly curved as $b$ approaches $b_{\rm max}$.  
The inset shows $h^{-2}$, which is  somewhat close to linear in $b$.  The curves drawn in Fig.\
\ref{h-obliq-k14+phi4} are from a modification of the $\varphi^4$ theory that was used for the NN model,
discussed in the section that follows.

\subsection{$\varphi^4$ theory for oblique--oblique DWs in the LRD model}
Now consider adding further neighbors to the NN theory.  Go back to the expression (\ref{dot-theta})
for $\dot\theta$, but supposing that more than just NN interactions are included.  Assuming
a static situation with $\theta=0$, $\phi=\frac{\pi}{2}+\psi$,  and using approximations 
$\cos\phi \approx -\psi$ and $\sin\phi \approx 1-\frac{1}{2}\psi^2$,  one has 
%
%
\begin{align}
\sum_{k=1}^R \frac{D}{k^3} & \Big[ (-2+3\psi^2) k^2 \psi_{xx}
 -3\psi\left(1-\tfrac{1}{2}\psi^2 \right) \left\{ 2-k^2 \psi_x^2 \right\} \Big]
\nonumber \\
& +2K_1 \psi\left(1-\tfrac{1}{2}\psi^2 \right) +\mu B^y \psi \approx 0 .
\end{align}
Length is measured in units of $a$ and $R$ is an upper limit for the sums. 
There is a sum over $1/k^3$, which is convergent for $R\to\infty$,
and a sum over $1/k$, which diverges for $R\to\infty$, but exceedingly slowly as $R$ increases. 
The divergence indicates an inconsistency between Taylor expansion and incorporation of long-range
interactions.  For a finite chain,  and much narrower DW, the sums would at minimum need to be truncated at
half the system size or possibly the DW width.  For most locations $x$, the offending dipole term  
$-\sum_k (2D/k) \psi_{xx}$ is close to zero due to $\psi_{xx}$ being small, except at the shoulders 
of a DW.  The other offending dipole term $-\sum_k (3D/k) \psi \psi_x^2$ is of even higher order 
of smallness and can be dropped. 

To make progress, the sum of $1/k$ is truncated at a small value of $k$.   Let the sums for a finite 
system be
\be
\zeta_1 = \sum_{k=1}^R \frac{1}{k}, \quad
\zeta_3 = \sum_{k=1}^{\infty} \frac{1}{k^3}.
\ee
Now I keep $\psi_{xx}$ as 3rd order but drop the higher order terms, such as $\psi_x^2$.  The result is
very similar to the NN $\varphi^4$ equation (\ref{phi4}), but with added constants $\zeta_1,\ \zeta_3$,
%
%
\be
\label{phi4L}
-2 \zeta_1 \psi_{xx} + (b-b_{\rm max}) \psi + \tfrac{1}{2} b_{\rm max} \psi^3 = 0,
\ee
where the limiting field for oblique states is now
\be
b_{\rm max}= 2(3\zeta_3-k_1).
\ee
This is the extension of the $\varphi^4$ theory to the LRD model.

\begin{figure}
\includegraphics[width=\figwidth,angle=0]{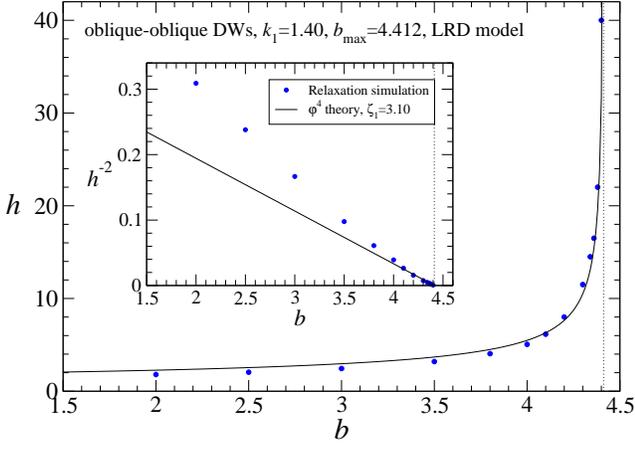}
\caption{\label{h-obliq-k14+phi4} For the LRD model with $k_1=1.40$, a comparison of the oblique-oblique
DW half-widths from relaxation simulations with those from the $\varphi^4$ approximate theory, Eq.\ (\ref{phi4w}).
The maximum magnetic field $b_{\rm max}=4.412$ for oblique states is shown by a dotted vertical line.
While the LRD value $\zeta_3 = 1.2020...$ was used, the other sum parameter was set using 12 neighbors,
$\zeta_1 = \sum_{k=1}^{12} \frac{1}{k} \approx 3.10$, which fits the limiting values as $b\to b_{\rm max}$,
but less-so for small $b$ where the DWs are very narrow.}
\end{figure}

The sum $\zeta_1$ produces a rescaling of the DW width. It implies that the hyperbolic tangent solution 
works here for oblique--oblique DWs,  
\be
\label{phi4-psi}
\psi = C \tanh \beta x, \quad
\ee
Eq.\ (\ref{phi4L}) is satisfied and gives a soliton solution for the constant values being
\be
C = \sqrt{\frac{2(b_{\rm max}-b)}{b_{\rm max}}}, \hskip 0.5in
\beta =  \sqrt{\frac{b_{\rm max}-b}{4\zeta_1}}.
\ee
That gives the half-width, 
\be
\label{phi4w}
h = \beta^{-1} = \sqrt{\frac{4\zeta_1}{b_{\rm max}-b}}.
\ee
As $\zeta_1$ is not exactly defined, the width has some uncertainty.  But the amplitude of this 
DW state is correct, as it connects exactly to the oblique states' angles.  which do not depend 
on $\zeta_1$.

\begin{figure}
\includegraphics[width=\figwidth,angle=0]{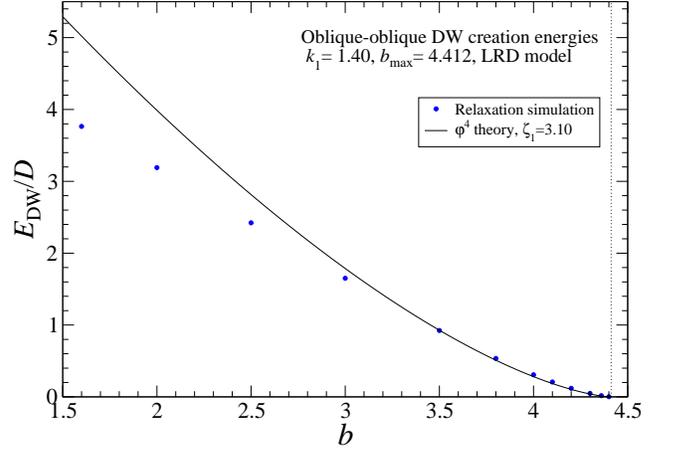}
\caption{\label{E-lrd-k14} For the LRD model, DW creation energy between oblique states for anisotropy
$k_1=1.40$, as a function of transverse applied field $b$.  Points are relaxation simulation data while the
curves are the approximate $\varphi^4$ theory result, Eq.\ (\ref{EDWL}),
using $\zeta_1=\sum_{k=1}^{12} \frac{1}{k} \approx 3.10$ . The simulated DW creation energy is the
difference in energies of the system with a DW present minus that in a relaxed oblique state.}
\end{figure}

The energy can be estimated, starting from a continuum expansion of the Hamiltonian in Eq.\ (\ref{Hphitheta}).  
The dimensionless energy density $u$ for static configurations, with approximate accounting of
LRD interactions, is
\begin{align}
u & = \sum_{k=1}^R \frac{1}{k^3} \big[ (1-3\cos^2\phi)\left(1-\tfrac{1}{2} k^2 \phi_x^2\right) \nonumber \\
& +\tfrac{3}{2} \left(\sin\phi\cos\phi\right) k^2 \phi_{xx} \big] -k_1\sin^2\phi-b\sin\phi .
\end{align}
Noting that sums $\zeta_1$ and $\zeta_3$ appear, doing an integration by parts for the total energy, and 
changing to the variable $\psi$ leads to the equivalent energy density [compare (\ref{upsi})],
\begin{align}
u  = & -2\zeta_3+\zeta_1 (1-\tfrac{3}{2}\sin^2\psi)\psi_x^2 \nonumber \\
& +\left[\left(3\zeta_3-k_1\right)\cos\psi-b\right]\cos\psi .
\end{align}
Now use the quadratic approximation for $\cos\psi$, and drop anything over quartic order of smallness,
which leads to a $\varphi^4$ energy density,
\be
u \approx u_{y\text{-par}} +\zeta_1\psi_x^2 -A \psi^2 + \tfrac{1}{2} B \psi^4,
\ee
where $u_{y\text{-par}}$, $A$, and $B$ are as defined in Sec.\ \ref{ssphi4}.
Then, the calculation of the DW creation energy is only slightly modified from that
for the NN model by the presence of $\zeta_1$ and $\zeta_3$, and the result is
\begin{align}
\label{EDWL}
\frac{E_{\rm DW}}{D} &=  \int_{-\infty}^{+\infty} AC^2 \sech^4\beta x\ dx
= \frac{8\sqrt{\zeta_1}}{3} \frac{(b_{\rm max}-b)^{3/2}}{b_{\rm max}}.
\end{align}
The sum $\zeta_1$ appears as a length rescaling that raises the energy, while $\zeta_3$ only increases 
$b_{\rm max}$, both relative to that for the NN model.

To give some idea whether this works, the half-width can be compared for the simulations
versus the $\varphi^4$ theory, as in Fig.\ \ref{h-obliq-k14+phi4}.  A similar comparison is
exhibited in Fig.\ \ref{E-lrd-k14} for the DW creation energy.  I used the infinite sum for 
$\zeta_3 = 1.2020...$, but twelve terms in $\zeta_1 \approx 3.10$.  This choice makes a good fit to
the energy especially as $b\to b_{\rm max}$, while overestimating the energy otherwise.  
This choice also fits the width well
as $b\to b_{\rm max}$ where the continuum approach is most reliable.  Then, the $\varphi^4$ theory 
slightly overestimates $h$ for $b$ farther from $b_{\rm max}$.  Also, the simulation
data for $h^{-2}(b)$ is a curve, while the $\varphi^4$ theory with fixed $\zeta_1$ can only give
a straight line.  Having the $\varphi^4$ theory slightly overestimate the creation energy is assumed, 
because the DW in the relaxation scheme can make slight changes relative to the approximate $\varphi^4$ 
shape to gain an energetic advantage.  This theory might be improved by a self-consistent calculation 
of the $\psi$-field together with an effective finite-range dipolar field that is a simple function 
related to the DW profile.
%

\subsection{DWs connecting $y$-alt to $y$-alt states in the LRD model}
It is interesting to consider $(k_1,b)$ parameters where uniform $y$-alt states are the most stable,
and investigate the possibility of DWs between them.  There are only two possible $y$-alt states,
and they look almost the same, except for whether the odd sites or the even sites have $\phi=+90^{\circ}$.
For reference, let the one with odd(even)  sites at $\phi=+90^{\circ}$ be called the odd(even) $y$-alt state.
A DW between $y$-alt states just connects these two over some distance. 

I started with an initial condition where each half of the system was given uniform and opposing spin directions
on the two sublattices, with inputted in-plane and out-of-plane angles, and then connecting the odd and 
even versions of these pseudo-$y$-alt states over one lattice constant.  A very small randomness was added to
the initial dipoles.  Then the system underwent the spin-pointing relaxation.  Different choices of initial 
angles could lead to different final configurations. 

%
%

\begin{figure}
\includegraphics[width=\figwidth,angle=0]{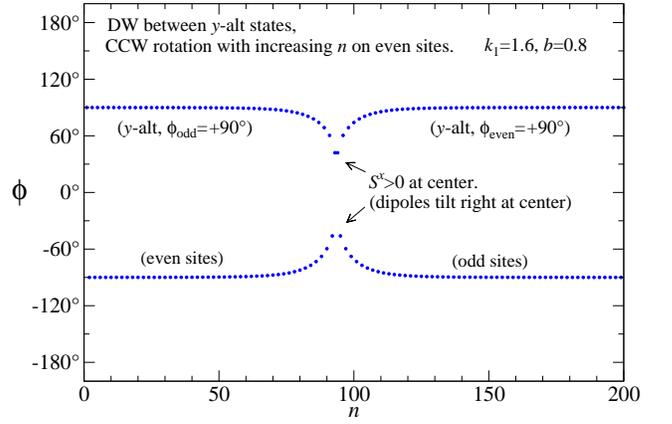}
\caption{\label{alt-kink-k16-b08-a} For the LRD model with parameters at point C in Fig.\ \ref{lrd-phases}), 
a DW with AFM local order connects the two $y$-alt states. It has a negative creation energy,
$E_{\rm DW} = -0.0329 D$, relative to $y$-alt.}
\end{figure}

\begin{figure}
\includegraphics[width=\figwidth,angle=0]{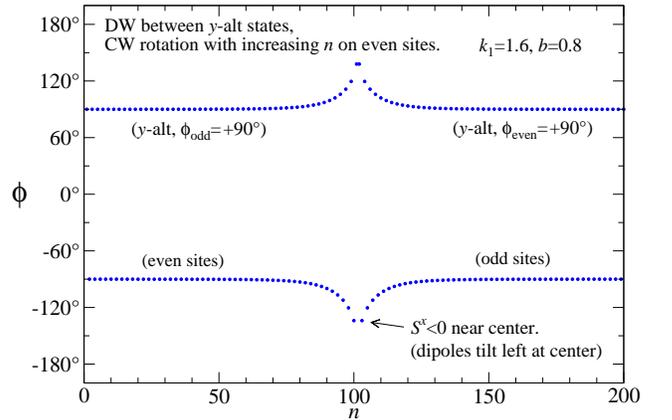}
\caption{\label{alt-kink-k16-b08-b} For the LRD model with parameters at point C in Fig.\ \ref{lrd-phases}), 
another DW with AFM local order connects the two $y$-alt states. It has a negative creation energy,
$E_{\rm DW} = -0.0329 D$, relative to $y$-alt.}
\end{figure}

For $k_1=1.6$ and $b=0.80$, examples of the results are shown in Figs.\ \ref{alt-kink-k16-b08-a} 
and \ref{alt-kink-k16-b08-b}.  At these parameters (point C in Fig.\ \ref{lrd-phases}, uniform $y$-alt and
oblique states are both possible but $y$-alt is lower in energy. The DWs show a significant width here.  In both 
examples, the odd $y$-alt state is on the left side, and connects to the even $y$-alt state on the right side.  
The difference, however, is that the odd dipoles can start at $90^{\circ}$ and then rotate to $-90^{\circ}$ 
either by going downward through $0^{\circ}$ (a CCW rotation as viewed from the $+x$-axis) or by going upward 
through $180^{\circ}$ (a CW rotation as viewed from the $+x$-axis).  This resembles two possible choices of helicity.
The even dipoles rotate in the opposite sense as the odd ones in both examples.  Another interesting aspect is 
that these DWs have the same {\em negative} creation energy, $E_{\rm DW} = -0.0329 D$, which is the
difference in energy of the DW state minus that of a uniform $y$-alt state. The negative value demonstrates 
that $y$-alt states are not the lowest energy configurations, and the DW state will be slightly preferred
at finite temperature.  In addition, the DW should be confined in the system, although any confinement potential 
seems to be very flat for the DW far from the chain ends. 
DWs of similar shape exist for $k_1=1.6$ at $b=0$, but with a positive creation energy, $E_{\rm DW} = 1.218 D$.


\begin{figure}
\includegraphics[width=\figwidth,angle=0]{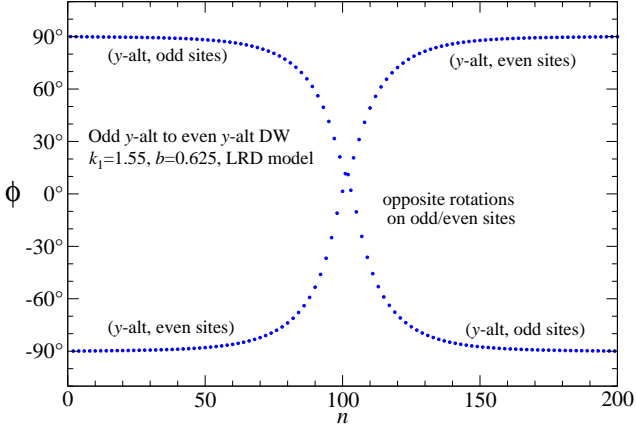}
\caption{\label{yalt-yalt-lrd-k155-b625} For the LRD model with $k_1=1.55, b=0.625$, a DW with AFM local order 
connects the two $y$-alt states.  These are parameters where uniform oblique and $y$-alt states have equal
energy per site.  It was found by accident while seeking oblique-to-$y$-alt DWs, which did not emerge.
It has a negative creation energy, $E_{\rm DW} = -0.1245 D$, relative to $y$-alt.}
\end{figure}

I chanced upon another negative-energy DW while trying to find oblique-to-$y$-alt DWs.  The parameters $k_1=1.55$ and 
$b=0.625$ make the energy per site for oblique and $y$-alt states the same.  Hence, I was using those values 
to look for an oblique-to-$y$-alt DW.  However, somehow the AFM ordering tendency is very strong and the 
whole system is dominated by $y$-alt.  The result is seen in Fig.\ \ref{yalt-yalt-lrd-k155-b625}, a very 
clean looking DW, separated on the odd/even sublattices, with creation energy $E_{\rm DW} = -0.1245 D$.  
%

\subsection{DWs connecting oblique to $y$-alt states in the LRD model}
In the last example, I mentioned doing simulations where per-site energies of uniform oblique and $y$-alt
states are the same, which should be expected to easily generate hybrid DWs between the two.  For the LRD
model, the energy densities are
\begin{align}
\frac{u_{\rm oblq}}{D} & = -2\zeta_R -\frac{(b/2)^2}{3\zeta_R-k_1}, \nonumber \\
\frac{u_{y\text{-alt}}}{D} & = -\tfrac{3}{4}\zeta_R-k_1.
\end{align}
The $y$-alt's energy is independent of $b$. When these are set equal for the same $k_1$, 
the required value of $b$ results, denoted as $b_{\rm oa}$ for {\em oblique-alternating}, 
\be
b_{\rm oa} = 2\sqrt{(3\zeta_R-k_1)(-\tfrac{5}{4}\zeta_R+k_1)}.
\ee
The formula applies to the limited region,
\be
k_1 > \tfrac{5}{4}\zeta_R \approx 1.5026 \quad \text{and} \quad
k_1 < \tfrac{17}{8}\zeta_R \approx 2.55437
\ee
The latter constraint comes from the using oblique stability limit, $b<\sqrt{2(3\zeta_R-k_1)}$.
The numerical values assume an infinite chain.

\begin{figure}
\includegraphics[width=\figwidth,angle=0]{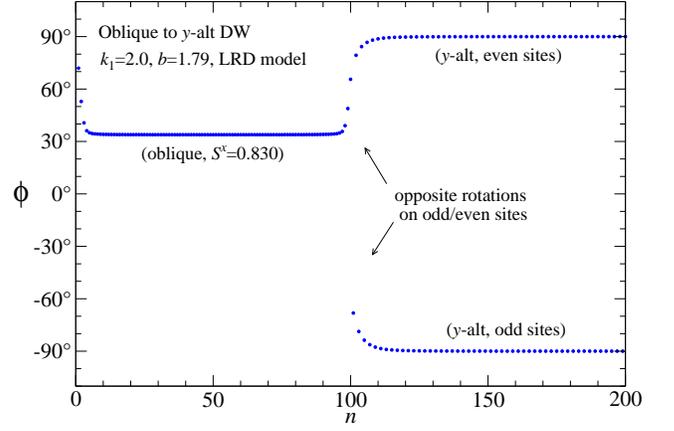}
\caption{\label{ob-yalt-lrd-k20-b179} For the LRD model with parameters at point D in Fig.\ \ref{lrd-phases}, 
a DW connects from an oblique state to a $y$-alt state. The DW creation energy relative to a fully oblique state  
is $E_{\rm DW} = -0.1963 D$.}
\end{figure}

Setting $k_1=2.0$ leads to the value $b_{\rm oa}\approx 1.79$, so these values were applied for another 
simulation.  The initial condition was an oblique state in half of the system abruptly connected to 
an alternating state in the other half. The relaxed result is shown in Fig.\ \ref{ob-yalt-lrd-k20-b179}.   
The hybrid $N=200$ system with a DW finds a tilting 
of the dipoles that barely lowers its energy to $E=-580.4088 D$, just below the pure states.  For comparison,
a $N=200$ system in oblique with minor end effects has energy $E_{\rm oblq}=-580.2125 D$, while in $y$-alt 
with no end effects it produces $E_{y\text{-alt}}=-579.4861D$.  These are not exactly equal due to end
effects. Then the creation energy is estimated as $E_{\rm DW} = E-E_{\rm oblq}= -0.1963 D$.

\begin{figure}
\includegraphics[width=\figwidth,angle=0]{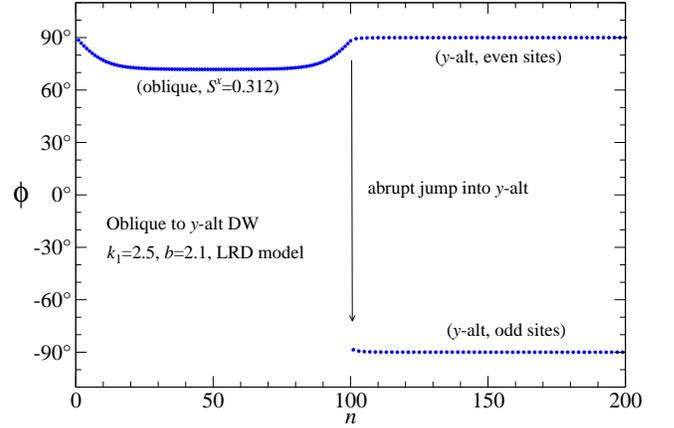}
\caption{\label{ob-yalt-lrd-k25-b21} For the LRD model with with parameters at point E in Fig.\ 
\ref{lrd-phases}, a DW connects from an oblique state to a $y$-alt state. The DW creation energy
is $E_{\rm DW} = -0.3027 D$, relative to a uniform oblique state.} 
\end{figure}

Another similar example is shown in Fig. \ref{ob-yalt-lrd-k25-b21}, for $k_1=2.5$  and $b=2.1$, 
where a uniform oblique state will be slightly lower in energy than a uniform $y$-alt state, due to
end effects. There is a very abrupt connection between the two states; the DW is especially narrow. 
Still, the system lowers its energy very slightly by forming this hybrid state: 
$E_{\rm DW} = E-E_{\rm oblq}=-0.3027 D$.
This type of state also can be found even if the oblique and $y$-alt energy densities are not too close.  
%

\subsection{DWs between $y$-par states?}
For $k_1 > 3.606$ and zero field, or smaller $k_1$ with moderate field, $y$-par states with $S^y=\pm 1$ 
coexist with $y$-alt, see Fig.\ \ref{lrd-phases}.  The $y$-par states are degenerate at zero field, implying 
it should be possible to form DWs between the two.  But these simulations tend to go to states with alternating 
dipoles, not always uniformly.  They do not tend to any smooth DW connecting $S^y=+1$ to $S^y=-1$.  If one does get 
a DW, it is abrupt and has a width of one lattice constant, that does not lend itself to a continuum analysis.

\section{Discussion and Conclusions}
A macrospin model is used here for a chain of magnetic islands oriented with their longer axes
transverse to the chain direction, also with a transverse applied magnetic field.  In earlier work,
the oblique, $y$-alt, and $y$-par uniform states were found to exist in partially overlapping regions 
of a phase diagram, see Fig.\ \ref{lrd-phases}.  In this work, static {\em non-uniform} 
states of this system were found.  

For anisotropy and field parameters $(k_1,b)$ where two types of uniform states are simultaneously stable, even 
with different energy densities), domain walls can exist that connect the two states. Due to the multiple  
choices of stable states, there are many types of DWs possible.  They have been found by numerical 
relaxation simulations that move the system towards local energy minima.  

For DWs connecting the two oblique states of opposite longitudinal magnetization, an approximate 
continuum theory leads to a description in terms of the static solitons of a $\varphi^4$ model.  
The DW itself is a transition region where the dipoles rotate over a potential 
energy barrier connecting the two energy minima, which are the uniform states. The top of the potential barrier, 
at the center of the DW, coincides with the dipoles being aligned with the magnetic field and aligned
with the long axes of the islands, minimizing those energies while maximizing their dipolar interaction energy.

The half-width $h$ and creation energies $E_{\rm DW}$ of oblique--oblique DWs are predicted analytically in the 
$\varphi^4$ model.  The analytic results compare well 
to numerical results found from the relaxation simulations (Figs.\ \ref{h-obliq-k14+phi4} and \ref{E-lrd-k14}), 
especially for applied field $b$ approaching the maximum ($b_{\rm max}$) that allows stable oblique states.  
That DW half-width increases with $(b_{\rm max}-b)^{-1/2}$ as $b\to b_{\rm max}$, wherein
the entire system will move into a $y$-parallel state with all dipoles pointing transverse to the chain. 
The creation energy of a DW, measured relative to a uniform oblique state, is found to increase 
proportional to $(b_{\rm max}-b)^{3/2}$.

The predictions from the $\varphi^4$ theory do not work if $b$ is far from $b_{\rm max}$,
such as in Figs.\ \ref{obliq-nn-af-k090-b0} and \ref{oblq-lrd-k14-b00} for $b=0$. In these cases,
the simulations indicate the existence of DWs between the oblique states where the 
DW itself exhibits AFM order, but the oblique states maintain ferromagnetic order.  
This surprising outcome must be a consequence of the NN dipole interactions. They have a strong tendency 
to induce site-by-site alternating order into the chain to minimize that energy contribution. 
The DW viewed on an individual sublattice appears to be close to hyperbolic tangent form.  A two-sublattice
analytic model that would include the possibility for AFM order, not in the scope of the current work,
may be able to verify that.  

The relaxation simulations also show that there are $y$-alt-to-$y$-alt DWs (such as Fig.\ \ref{alt-kink-k16-b08-a}) 
in which most of the system has AFM order, but modulated by a smooth spatial transition where the even sites 
rotate from $\phi_{\rm even}=-90^{\circ}$ to $\phi_{\rm even}=+90^{\circ}$ while the odd sites rotate 
in the opposite sense from  $\phi_{\rm odd}=+90^{\circ}$ to $\phi_{\rm odd}=-90^{\circ}$.  
Each sublattice has a smooth behavior that might be analyzed in a two-sublattice 
continuum limit.  Surprisingly, these DWs were found to have negative creation energy, relative to the energy 
of a uniform $y$-alt state.  Once again this is likely attributable to the strong influence of NN dipole 
interactions. It implies that a uniform $y$-alt {\em is not} the lowest energy configuration where it is stable 
in the $(k_1,b)$ diagram of Fig.\ \ref{lrd-phases}.  

Oblique-to-$y$-alt DWs were produced in the simulations (such as Fig.\ \ref{ob-yalt-lrd-k20-b179}), also 
with negative creation energy relative to the lower-energy (oblique) phase.  Then, the system could exist 
very stably in a state with ferromagnetic order separated from antiferromagnetic order by the DW.  It might 
be possible to analyze this situation using a two-sublattice continuum theory.  

The existence of the different DWs in this theoretical model depends particularly on the anisotropy strength 
and applied field, as well the assumed perfect dipole pair couplings. Any external perturbations could
modify the interactions and change the DW structure and stability.  Possibly, these DW configurations can be 
exploited as detectors of small external influences or applied field changes.

\end{document}